\documentclass[aps, prl, twocolumn, showpacs, superscriptaddress]{revtex4}
\usepackage[dvips]{graphicx}
\usepackage{amssymb}
\usepackage{amsmath}
\usepackage{color}

\begin{document}

\title{Mechanical Flip-Chip for Ultra-High Electron Mobility Devices }

\author{K. Bennaceur}
\author{B.A. Schmidt}
\author{S. Gaucher}
\affiliation{Department of Physics, McGill University, Montr\'eal, Quebec, H3A 2T8, Canada}
\author{D. Laroche}
\affiliation{Department of Physics, McGill University, Montr\'eal, Quebec, H3A 2T8, Canada}
\affiliation{Center for Integrated Nanotechnologies, Sandia National Laboratories, Albuquerque, NM 87185 USA}
\author{M.P.~Lilly}
\author{J.L Reno} 
\affiliation{Center for Integrated Nanotechnologies, Sandia National Laboratories, Albuquerque, NM 87185 USA}
\author{K.W. West}
\author{L.N. Pfeiffer}  
\affiliation{Department of Electrical Engineering, Princeton University, Princeton, NJ  08540 USA  }
\author{G. Gervais}
\affiliation{Department of Physics, McGill University, Montr\'eal, Quebec, H3A 2T8, Canada}

\begin{abstract}
Electrostatic gates are of paramount importance for the physics of devices based on high-mobility two-dimensional electron gas (2DEG) since they allow depletion of electrons in selected areas. This field-effect gating enables the fabrication of a wide range of devices such as, for example,  quantum point contacts (QPC), electron interferometers and quantum dots. To fabricate these gates, processing is usually performed on the 2DEG material, which is in many cases detrimental to its electron mobility. Here we propose an alternative process which does not require any processing of the 2DEG material other than for the ohmic contacts. This approach relies on processing a separate wafer that is then mechanically mounted on the 2DEG material in a flip-chip fashion. This technique proved successful to fabricate quantum point contacts on both GaAs/AlGaAs materials with both moderate and ultra-high electron mobility.
\end{abstract}

\maketitle

Electrostatic gates fabricated on the sample surface are routinely used to locally deplete the two-dimensional electron gases (2DEGs) in semiconductor heterostructures, in order to study the electronic behaviour in further confined geometries. The realization of the split gate in 1981 \cite{SplitGate} opened up the possibility to observe several new quantum effects in electron transport and as such generated a field of research on its own. Arguably the most famous use of split gates is the quantum point contact (QPC) \cite{QPC}, whereby a narrow constriction with a tuneable width comparable to the Fermi wavelength is fabricated on a 2DEG. This led to the observation of one-dimensional ballistic transport, a regime where the conductance $g$  is quantized in even steps of $2e^2/h$ as a function of the constriction width. The use of QPCs has generated an enormous body of work regarding ballistic quantum transport \cite{Beenakker} and its impact on mesoscopic physics has been enormous. Nowadays, very similar fabrication techniques are used to fabricate devices tailored for the study of  electronic transport in quantum dots \cite{Qdot}, electron interferometers \cite{Heiblum}, and phase-coherent mesoscopic circuits \cite{Glattli}.

When the electron mobility in a 2DEG surpasses $\rm{ \sim10^{5}~cm^2/V\cdot s}$, drastic quantum effects can arise in a magnetic field, such as the Fractional Quantum Hall Effect (FQHE). This counter-intuitive phenomenon involves the two-dimensional system acquiring {\it fractional} effective charges, quantum statistics and quantum numbers, all driven by electron-electron interactions.  This is in stark contrast with the Integer Quantum Hall Effect (IQHE) whose emergence does not involve any interactions and consequently is a much more robust phenomenon against disorder.  While detrimental to the electronic mobility of its 2DEG, split gates fabricated on high-mobility GaAs/AlGaAs heterostructures have led to important insights.  For instance,  shot noise \cite{e/3} measurements were used to  determine effective charges in FQHE circuits;  electron interferometry with FQHE quasiparticles are now performed in Fabry-Perot \cite{FPI}, and/or the Mach-Zehnder \cite{MZI} interferometers which are electronic equivalent to those used in optics. A serious drawback, however, is the processing required to fabricate these gates, which results in unwanted degradation of the electron mobility. This is particularly damaging for delicate many-body quantum states such as the 5/2 and 12/5 FQH states. These states are unusual in that their quantum statistics is believed to emanate from a non-Abelian lineage \cite{Ady}, but unfortunately their small energy many-body gap of $\sim$500~mK and 50~mK respectively, is affected by disorder of any kind. As such,  it has been difficult thus far to study these states in gated structures modulo a few {\it tour de force} experiments  performed on the 5/2 FQH state \cite{QPC5/2,Willett}.\\


In order to circumvent the problem of degradation of the electronic mobility,  we have developed a flip-chip technique where all processing steps required to fabricate the electrostatic gates are performed on a separate substrate that is then mounted on the surface of the 2DEG.  This approach has previously been used to fabricate resonators on 2DEGs \cite{Reulet} as well as for the  coupling of high-frequency lines \cite{FCHF}; however, to our knowledge, split gates on high-mobility structures have never been reported using a flip-chip technique. This technique offers several advantages:\\

 - {\it It avoids contamination by chemicals  during the fabrication process}. Usually, the gate fabrication process uses several steps in which polymer resists are deposited on the surface of the 2DEG and chemicals are used to develop and remove these resists. Even when an appropriate solvent is used, these resist residues are hard to remove and doing so often involves a cleaning process that can damage the 2DEG ({\it e.g.} an oxygen plasma). Resist residues can also trap charges and generate undesirable fluctuations in the density of the 2DEG.\\

 - {\it It avoids degradation of the electron mobility during lithography}. Radiation damage can arise from heating, electrostatic charging, ionization damages (radiolysis), displacement damage, sputtering and hydrocarbon contamination. In our approach, the 2DEG is never exposed to any electron beam or optical lithography.\\

 - {\it It avoids additional strain induced by differential thermal contractions}. At room temperature,  GaAs has a thermal expansion coefficient of $\alpha_L=5.8\times 10^{-6} \, \rm{K}^{-1}$. Electrostatic gates are typically made with a few nanometers of Ti ($\alpha_L=8.6\times10^{-6} \, \rm{K}^{-1}$) and/or Cr ($\alpha_L=4.9\times10^{-6} \, \rm{K}^{-1}$) adhesion layer, and most often with tens of nanometers of gold ($\alpha_L=\rm{14\times10^{-6} \,K^{-1}}$) or aluminium ($\alpha_L=23.1\times10^{-6} \, \rm{K}^{-1}$). These coefficients do not vary much from $\sim$300~K down to 150~K and eventually become negligible at low temperatures. For a typical 2~mm wide gate made out of gold,  the gate will shrink by approximately $4 \, \rm{\mu m}$ during the first 150~K, whereas  GaAs will shrink by less than $\rm2~\mu m$. This is likely to induce strain in the material and to affect the electron mobility at low temperatures, particularly in the case of  a narrow constriction with a width of only  500~nm.\\

  - {\it It makes it possible to re-use the 2DEG material}. The technique is non-destructive and enables the re-use of the 2DEG material at will. In addition, for the highest mobility GaAs/AlGaAs 2DEGs with mobility in excess of $30\times 10^{6} \, \rm{cm^2/V\cdot s}$, it is known that different parts of the wafer may have different mobilities as well as fluctuations in the electron density. Since several types of gated devices have active areas of a few $\rm{\mu m^2}$ only, should the need arise, our approach allows for the device to easily be remounted on a slightly different part of the wafer.  This technique also allows one to swap gates so as to measure different devices (or designs) on the exact same piece of material. Finally, it avoids wasting precious material during low-yield processes, which is common when fabricating complex devices.\\

\begin{widetext}

\begin{figure}[t]
  \includegraphics[width=18cm]{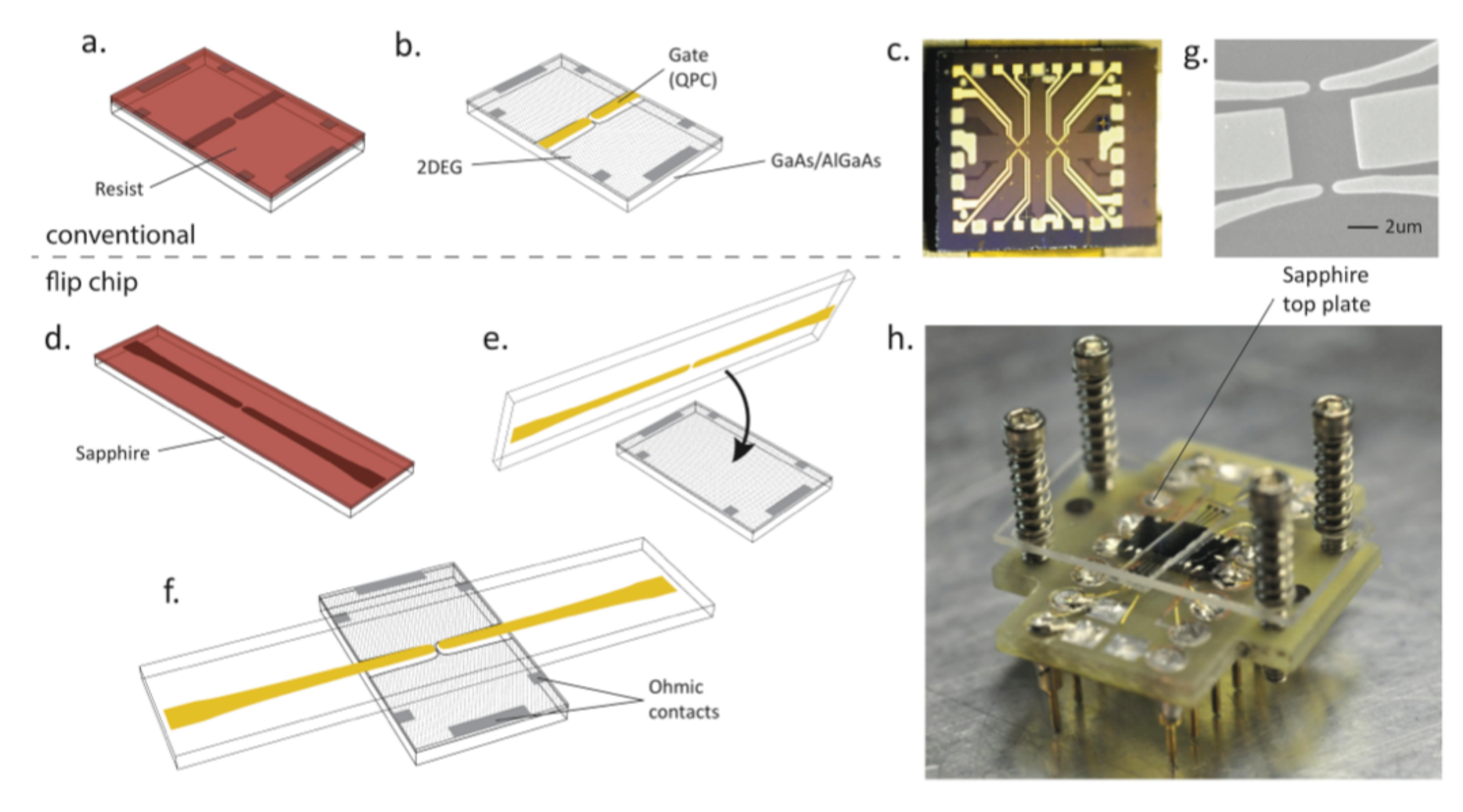}
  \caption{Top panels: conventional process. In {\bf a.} and {\bf b.} the lithography is performed directly on the 2DEG wafer. A photograph of a typical device is shown in {\bf c}.  Bottom panels: the flip-chip process. {\bf d.} e-beam lithography is performed on a sapphire substrate and then {\bf e.} the metallic gates are brought into proximity of the 2DEG by a flip-chip process. The final assembly is shown in {\bf f}.  A photograph of the central part of the gates as well of the whole device is shown in {\bf g.} and {\bf h.}, respectively.}
  \label{diagram}
\end{figure}

\end{widetext}

\begin{figure}[!h]
  \includegraphics[width=9cm]{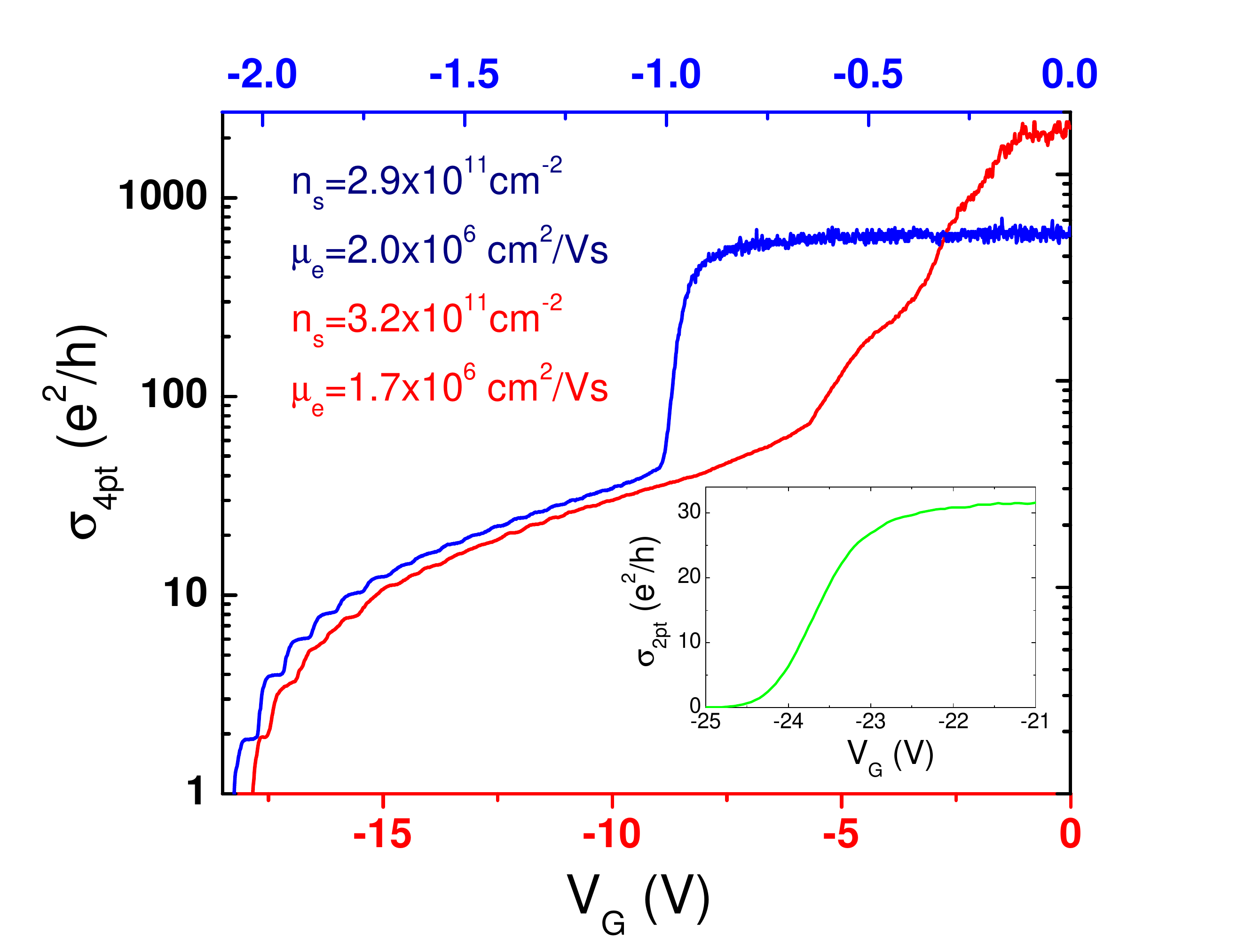}
  \caption{Conductance across the 2DEG as a function of gate voltage (blue, top axis) applied on the conventional QPC, and on the FCQPC (red, bottom axis, sample A). The inset shows a pinch off of a FCQPC at 4~K fabricated on an ultra-high mobility 2DEG wafer (sample B). }
  \label{QPC}
\end{figure}


In Fig.\ref{diagram} we show the conventional (panels {\bf a-c}) and flip-chip (panels {\bf d-h})  fabrication processes of a QPC. In the traditional approach, the gates are fabricated directly on top of the 2DEG material using a combination of e-beam and optical lithography.  In contrast, the flip-chip technique uses a sapphire plate where all the processing is performed through a conventional
$e$-beam lithography process, with 50 nm of chromium first deposited to allow charges evacuation during the e-beam exposure followed by a spinning of MAA/PMMA resist. After exposure and development, 5/150 $nm$ Ti/Au is deposited by $e$-beam evaporation followed by lift-off and chromium etching. Extreme care is taken to obtain a near perfect lift-off so as to prevent residual metal standing up along the edge, which could cause gate leakage or introduce an undesirable buffer space between the flip-chip gates and the 2DEG. A layer of aluminum oxide (30 to 100~nm thick) is then deposited to further prevent potential gate leakages. The ohmic contacts are fabricated directly on the GaAs/AlGaAs either by indium diffusion, or by evaporation of Ge/Au/Ni/Au using shadow mask to avoid lithography. The 2DEG wafer is placed on a sample holder and the ohmic contacts are connected to the contacts of the holder by indium-soldered gold wires. Then, the flip-chip is placed on top of the 2DEG together with an additional sapphire top plate, which is itself held in place by four BeCu springs  that apply very gentle mechanical pressure. The optical interference fringes between the top plate and the flip-chip allow a fine tuning in the alignment of the flip-chip device. All fabrication and assembly steps are performed in a class-100 cleanroom to avoid contamination of the critical interfacing surfaces. \\

We have fabricated both $1$ $\mu m$ long QPC and Fabry-P\'erot gates on sapphire and tested the technique on GaAs/AlGaAs wafers grown at Sandia National Laboratories (Sample A) as well as  Princeton University (Sample B). We have chosen to compare the flip-chip devices with the best conventional QPC fabricated  out of 20; in many cases, the conductance quantization were not as good as for the flip-chip devices. For the flip-chips mounted on sample A, several QPCs showed pinch-off with a success rate of $\sim$80\%. In total, more than 20 devices were tested and  a pinch-off voltage ranging from -6~V to -40~V was determined. For comparison, conventional QPCs (CQPCs) made on a similar wafer to sample A showed a pinch-off voltage ranging from -1.7 to -3 V, and thus had a superior gate efficiency by a factor ranging from $\sim$3 to 13. We attribute the lower gate efficiency  of the FCQPC to the `air' (vacuum) gap between the flip-chip and the 2DEG. A rough calculation that assumes a pinch-off voltage scaling linearly with the distance between gates and the 2DEGs estimates this gap to be in the range from 50 to 200~nm. This gap most likely occurs because the contact area between the 2DEG and the flip-chip gates is over  $3.6\times 2 ~mm^2$ in area and  not perfectly flat; this could easily be improved by reducing the contact area between the two mechanical parts of the device. \\

\begin{figure}[!h]
  \includegraphics[width=8cm]{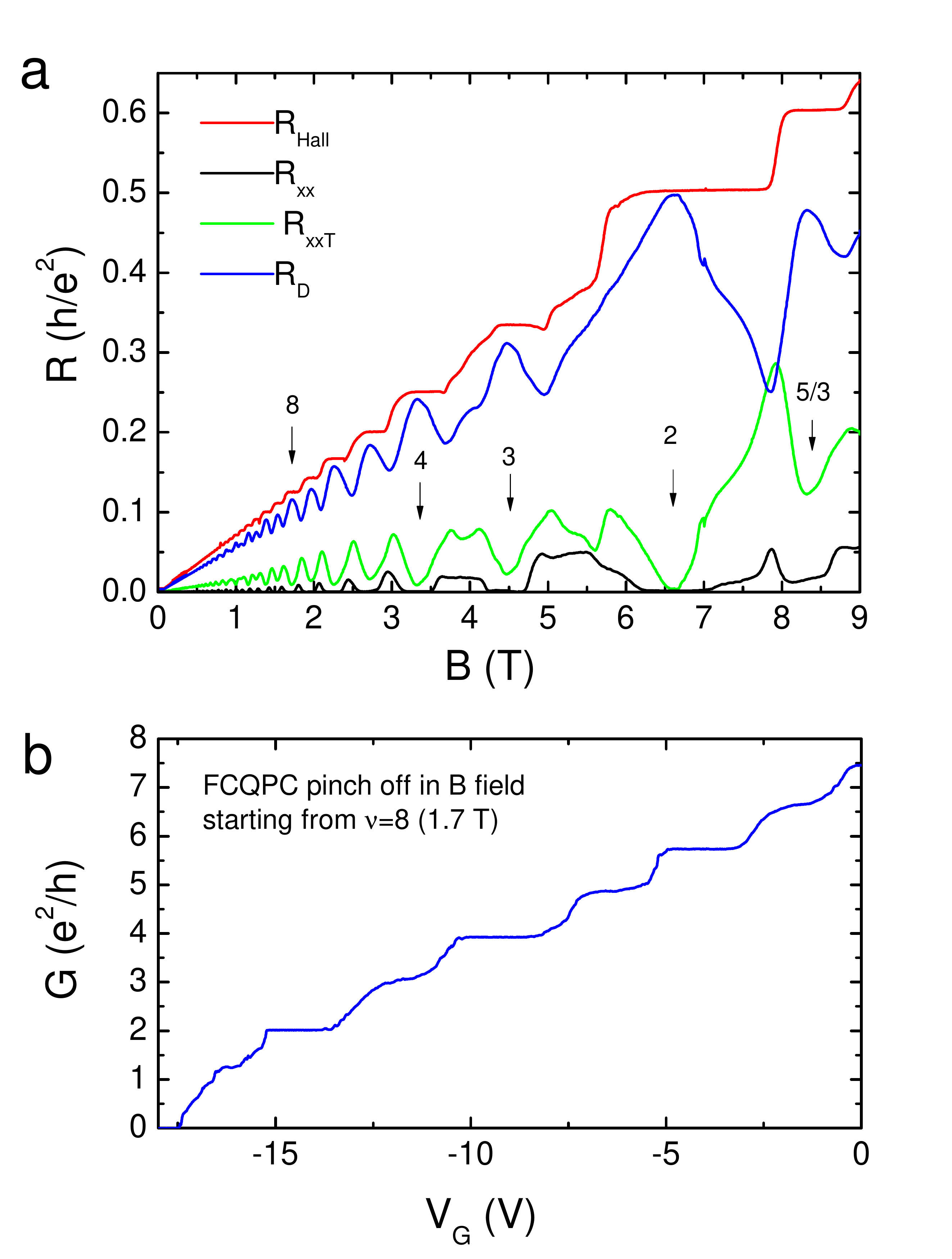}
  \caption{{\bf a.} Hall resistance ($R_H$), longitudinal resistance outside ($R_{xx}$) and through ($R_{xxT})$ the FCQPC, as well as the diagonal resistance $R_D$ (mixing of $R_H$ and $R_{xxT}$)  versus the magnetic field measured on sample A.  Clear QHE and FQHE features (such as the 5/3 FQH state) can be observed. {\bf b.} Diagonal conductance across the QPC at 1.7~T magnetic field versus gate voltage.}
  \label{Resistance}
\end{figure}

Figure \ref{QPC} provides a typical example of the conductivity versus pinch-off gate voltage on a logarithmic scale for both a flip-chip QPC (red) and a conventional QPC (blue). The data were taken at 25~mK using a similar 2DEG GaAs/AlGaAS heterostructure with a moderate mobility of $2\times 10^{6}$~$\rm cm^2/V\cdot s$ (although not coming from the exact same wafer). For both devices, the quantization of the conductance is clearly observed. While the CQPC shows better-defined and more precise conductance plateaus (ballistic behaviour), this likely arises  from the shape of the QPCs that were not identical, and because of  both the air gap present in the FCQPC and shallowness of the 2DEG modifying the gate-induced electric field profile.  We also note that in the metallic regime at conductance below $\sim50$ $e^2/h$,  the slopes of the FCQPC and CQPC are strikingly similar.  Importantly, we have not observed a pronounced hysteresis in the pinch-off curve of the FCQPC whereas this undesirable behaviour is often observed in conventional gated devices. Finally, in the inset of Figure \ref{QPC} the conductance of a flip-chip QPC integrated with an ultra-high mobility ($\mu\sim 1.0 \times 10^7 \, \rm{cm^2/V\cdot s}$) GaAs/AlGaAs 2DEG and measured at 4~K is shown. Albeit with a lower gate efficiency, this data demonstrates that the flip-chip process can be integrated with the highest mobility 2DEG materials. \\

As further confirmation of the quality and efficiency of the flip-chip devices, they were characterized in the quantum Hall regime at very low temperatures in a magnetic field up to 9~T. Figure \ref{Resistance}{\bf a} shows the measured resistance in various configurations within sample A. The Hall resistance $R_H$ measured outside of the interferometer (red trace) shows good quantization of the IQH plateaus and the 5/3 FQH plateau, as expected for a sample in this range of mobility. Due to the geometry of our design, it is not possible to directly measure $R_H$ under the interferometer; rather, we measure the diagonal resistance $R_D$ (blue trace), which includes a large longitudinal contribution. Also shown are $R_{xx}$ outside of the interferometer and $R_{xxT}$ through the interferometer; again geometry plays a role, since the contacts outside are much closer together and the measured voltage is correspondingly smaller. The fact that $R_{xxT}$ does not reach zero in all of the states and that $R_{xx}$ does may be due to the flip-chip blocking light from the LED used during the cooling process or to the non-ohmicity of particular contacts. Further investigation is required to clarify this. \\

In Figure \ref{Resistance}{\bf b}, the conductance is shown as a function of gate voltage at a magnetic field of 1.7~T, near filling factor $\nu=8$. Plateaus occur for each integer multiple of $e^2/h$, with the exception of the first plateau. The quantization for the lowest conductivity plateaus (2, 3, 4) is excellent, but degrades for higher plateaus as the contribution of the longitudinal resistance in series with the QPC becomes significant. Along with the data in Figure \ref{Resistance}{\bf a}, these results clearly show that the device can operate in a magnetic field and is mechanically stable up to 9~T at 20~mK.\\

In conclusion, we have demonstrated successful fabrication of split-gate devices on GaAs/AlGaAs 2DEG heterostructures using a flip-chip technique which avoids performing any potentially damaging fabrication steps directly on the ultra-high mobility wafer. This process has a yield on par with traditional fabrication techniques, and results in robust devices that can be measured at dilution refrigerator temperatures and in high magnetic fields. Benchmark measurements show that the samples perform as expected; we demonstrate pinch-off and quantized conductance both at zero field and in the quantum Hall regime. These devices have great potential for future studies of sensitive quantum Hall states such as interferometric measurements with the 5/2 and 12/5 FQHE that occur only in pristine ultra-high mobility devices. \\



We thank the clean room  facilities at McGill, {\'E}cole Polytechnique de Montr\'eal, and Universit\'e de Sherbrooke for providing us access to their complexes. This work was funded by NSERC (Canada), FRQNT (Qu\'ebec) and  the Canadian Institute for Advanced Research (CIFAR). Part of this work was performed at the Center for Integrated Nanotechnologies, a U.S. DOE, Office of Basic Energy Sciences, user facility. Sandia National Laboratories is a multiprogram laboratory managed and operated by Sandia Corporation, a wholly owned subsidiary of Lockheed Martin Corporation, for the U.S. DOE National Nuclear Security Administration under contract DE-AC04-94AL85000. The work at Princeton was partially funded by the Gordon and Betty Moore Foundation as well as the NSF MRSEC Program through the Princeton Center for Complex Materials (DMR-0819860).  We also thank R. Talbot, R. Gagnon, and J. Smeros for technical assistance. All data, analysis details and material recipes presented in this work are available upon request to G.G.

\end{document}